\documentstyle[epsfig,aps]{revtex} 
\begin{document}
\newcommand{\p}{\partial}
\newcommand{\ls}{\left(}
\newcommand{\rs}{\right)}
\newcommand{\beq}{\begin{equation}}
\newcommand{\eeq}{\end{equation}}
\newcommand{\beqa}{\begin{eqnarray}}
\newcommand{\eeqa}{\end{eqnarray}}
\newcommand{\bdm}{\begin{displaymath}}
\newcommand{\edm}{\end{displaymath}}

\twocolumn[\hsize\textwidth\columnwidth\hsize
           \csname @twocolumnfalse\endcsname]
\title{Softening of the nuclear 
equation-of-state by kinetic non-equilibrium in heavy ion collision
}
\author{
C. Fuchs$^{1}$ and T. Gaitanos$^{2}$}
\address{
$^{1}$Institut f\"ur Theoretische Physik der Universit\"at T\"ubingen, 
Auf der Morgenstelle 14, D-72076 T\"ubingen, Germany\\
$^{2}$ Sektion Physik der Universit\"at M\"unchen, 
Am Coulombwall 1, D-85748 Garching, Germany\\
}
\maketitle  
\begin{abstract}
Highly compressed nuclear matter created in relativistic heavy 
collisions is to large extent governed by local non-equilibrium.
As an idealised scenario colliding nuclear matter 
configurations are studied 
with an effective in-medium interaction based on the microscopic 
DBHF model. It is found that on top of the repulsive 
momentum dependence of the nuclear forces 
kinetic non-equilibrium leads to an effective softening 
of the equation of state as compared to ground state 
nuclear matter. The separation of phase space which is the 
basic feature of such anisotropic configurations has thereby a 
similar influence as the introduction of a virtual new degree 
of freedom. 
\end{abstract}
One major goal of relativistic heavy ion physics is 
to explore the behaviour of the nuclear equation-of-state (EOS) far 
away from saturation, i.e. at high densities and non-zero 
temperature. Over the last three decades a large variety of observables 
has been investigated both, from the experimental and theoretical 
side motivated by the search for the nuclear EOS. 
The collective particle flow is thereby intimately connected to the 
dynamics during the compressed high 
density phase of such reactions \cite{stoecker86,ritter97}. 
E.g., the elliptic flow which develops in the early 
compression phase is thought to be a suitable observable 
to extract information on the EOS \cite{eflow}. But also the 
production of strange particles is a good probe to study dense 
matter \cite{AiKo85}. Recent precision measurements of the $K^+$ production 
at SIS energies strongly support the scenario of a relatively soft 
EOS \cite{sturm00,fuchs00}. 

The temporal evolution of the collision from a highly anisotropic 
initial configuration in phase space to an -- at least 
partially -- equilibrated final 
configuration is successfully described by microscopic 
transport models like BUU \cite{Cassing99} or QMD \cite{Ai91}. 
In this type of models the nuclear mean field is usually 
based an phenomenological parameterisations 
\cite{Cassing99,Ai91,welke88,dani00}. 
Such parameterisations allow 
different extrapolations to high densities, subsummized 
by referring to a 'hard' or a 'soft' equation-of-state \cite{stoecker86},  
which can be tested in heavy ion collisions. 
A fundamental question is, however, if such a procedure is sufficient to 
to extract well defined information 
on the EOS of equilibrated nuclear matter (NM) from 
heavy ion collisions. The mean field used in 
transport simulations refers to {\em locally equilibrated nuclear 
matter}. Anisotropy effects of the local phase space are usually 
neglected for the density dependent (or local) part 
of the mean field although 
it is clear that they should be included from a theoretical point 
of view \cite{btm90,gmat1,sehn96,lca}. Such a 
treatment appears to be justified if non-equilibrium effects are 
small, have a short lifetime, or if the mean field is not 
much affected. As we will discuss in the present work 
none of this three conditions is fulfilled in relativistic heavy 
ion reactions in the SIS energy range: 
1. The initial phase space distribution in the 
participant zone is that of two currents of 
nuclear matter colliding with beam 
velocity. The local momentum distributions of colliding nuclear matter 
configurations, called CNM in the following, are given by 
two Fermi ellipsoids, i.e. two boosted Fermi spheres separated 
by a relative velocity \cite{sehn96}. 
2. The relaxation time needed by the system to equilibrate 
coincides more or less with the high density phase of the reaction. 
Hence, anisotropy effects are present all 
over the compression phase where one essentially intends to 
study the EOS at high densities. 
Experimental evidence for incomplete equilibration 
even in central collisions at SIS energies 
has recently been reported in \cite{rami99}. 
3. The impact of phase space anisotropies on the nuclear EOS is 
large and leads to a considerable softening of the 
effective EOS seen in heavy ion collisions.

To obtain a more quantitative measure for the size and relevant time scales 
for phase space anisotropies in Fig. 1 the time evolution of the 
quadrupole moment of the energy-momentum-tensor
\beq
Q_{zz} = \frac{2T^{33} - T^{11} -T^{22}}{T^{33} + T^{11} + T^{22}}
\eeq
at the collision centre of central (b=0 fm) $Au+Au$ 
reactions at two typical beam energies, 
i.e. 0.6 A.GeV and 6.0 A.GeV, respectively, is compared to the 
time evolution of the corresponding baryon density. The quantities 
were obtained in relativistic BUU calculations \cite{lca}. 
$Q_{zz}$ is a measure for the 
anisotropy in beam ($z$) direction. At 6.0 A.GeV $Q_{zz}$ reaches 
already the asymptotic value of 2 where the Fermi momentum can be 
neglected compared to the beam velocity. Fig. 1 demonstrates 
that the relaxation time to 
reach equilibrium configurations ($Q_{zz} \simeq 0$) coincides 
more or less with the high density phase, independent of the 
beam energy. Therefore non-equilibrium effects should be 
taken into account on the level of the effective in-medium 
interaction which means to determine the mean field used in 
transport calculations consistently for 
colliding nuclear matter.

\begin{figure}[h]
\unitlength1cm
\begin{picture}(8.,5.3)
\put(-0.5,0.0){\makebox{\epsfig{file=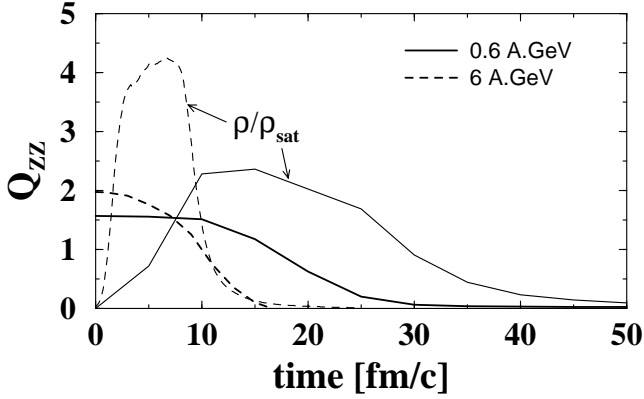,width=8.5cm}}}
\end{picture}
\caption{Time evolution of the quadrupole moment $Q_{zz}$ 
(bold lines) at the collision centre in central $Au+Au$ reactions 
at 0.6 and 6.0 A.GeV. $Q_{zz}$ is a measure for the 
anisotropy of the local momentum space during the reaction. 
For comparison also the 
central densities (in units of $\rho_{\rm sat}$) are shown.}
\label{Fig1}
\end{figure}

In the present work we want to study the impact of such non-equilibrium 
effects on the nuclear EOS probed by heavy ion collisions. The 
configurations are idealised by colliding nuclear matter which 
describes well the local phase space in the participant zone 
of the collisions. CNM also provides a smooth transition to 
equilibrated configurations which evolve in the later stages of the 
reactions \cite{essler97}. The effective interaction used in 
the present studies is based on the microscopic 
relativistic Dirac-Brueckner-Hartree-Fock (DBHF) 
approach \cite{dbhf,boelting99} which turned out to be 
quite successful in the description of 
nuclear matter bulk properties. 
Here the nucleon-nucleon interaction 
is fixed by free NN-scattering and the medium correlations 
contained in the self-consistent 
summation of the ladder diagrams of the Bethe-Salpeter equation 
determine the behaviour of the nuclear EOS. 
DBHF forces have been extensively tested at 
SIS energies below 1 A.GeV, and 
a general agreement with corresponding flow data has been observed 
\cite{dani00,lca,fopi95}. 
The mean field used in the present work is based on recent 
DBHF calculations \cite{boelting99} with nuclear matter 
saturation properties (Bonn A potential) 
of $\rho_{\rm sat}= 0.185~{\rm fm}^{-3},~E^{\rm bind} = -16.15$ MeV and 
a compression modulus of $K= 230$ MeV. 
The solution of the full DBHF problem for colliding nuclear 
matter, i.e. two-Fermi-ellipsoid configurations, is, however, 
still an unresolved problem and thus we determine the effective 
interaction in CNM as discussed in \cite{sehn96}. 
By this procedure the density and momentum dependence 
of the DBHF self-energies are extrapolated to the CNM configurations 
and the mean field is constructed by a superposition of the 
contributions from the two currents. However, already this 
approximation invokes a self-consistency problem for 
the CNM configuration 
\beq
\Theta_{12} = \Theta_{1} +  \Theta_{2} - \Theta_{1}\cdot \Theta_{2}
\label{conf}
\eeq
composed by the two currents 
$ \Theta_{i} =  \Theta (\mu^*(k_{F_{i}}) - k_{\nu}^* u_{i}^\nu )$. 
The effective mass $ M^*=M+{\Sigma}_{S_{12}}$,  
the scalar density $\varrho_{s_{12}} = <M^* /E^* >_{12}$ and 
the configuration (\ref{conf}) are coupled by non-linear equations. 
$ \Theta$ is the step function, $k_{F_{i}}$ are the Fermi 
momenta and $ u_{i}^\nu = (\gamma_i, \gamma_i {\bf u}_i)$ 
are the streaming velocities of the two 
subsystem currents. The last term in eq. (\ref{conf}) ensures that the 
Pauli principle is fulfilled for small velocities where 
the two ellipsoids might overlap. For details see \cite{sehn96}.  
\beq
<X >_{12} = \frac{\kappa}{(2\pi)^3} 
\int d^3 {\bf k} X(k) \Theta_{12} (k)
\label{confint}
\eeq
denotes the summation over all occupied states. 
In spin-isospin saturated nuclear 
matter the phase space occupancy factor is $\kappa =4$. 
The energy momentum 
tensor in CNM is given by \cite{sehn96} 
\beq
        T^{\mu\nu}_{12} = 
       < k^{\mu} k^{\ast\nu} / E^{\ast} >_{12} 
        - V^{\mu\nu}_{12}
        - \frac{1}{2} g^{\mu\nu} \left\{ \overline{\Sigma}_{S_{12}}\,
        \varrho_{s_{12}}
        - V^{\lambda}_{12 \, \lambda}   \right\}
\label{tfull}
\eeq    
with the scalar contribution 
\beq
\overline{\Sigma}_{S_{12}} =  
< \Sigma_{S_{12}} M^* / E^{\ast} >_{12} / \varrho_{s_{12}}
\label{scalar}
\eeq
 and 
the terms arising from the vector field 
\beq
V^{\mu\nu}_{12} = < \Sigma_{12}^{\mu} k^{\ast\nu} / E^{\ast} >_{12}
\quad .
\label{vector}
\eeq
The scalar self-energy $\overline{\Sigma}_{S_{12}}$ enters in a Hartree 
form averaged over the explicit momentum dependence into 
(\ref{tfull}). In an analogous 
way a Hartree vector self-energy 
$\overline{\Sigma}_{12}^{\mu}= 
V^{\mu\lambda}_{12} j_{12\lambda} / j_{12}^2$ can be defined. 
The full self-energy has the form $
 \overline{\Sigma}_{12} (k_{F_{1}},k_{F_{2}}, u_1, u_2) 
= \overline{\Sigma}_{S_{12}}
+ \gamma_\mu \overline{\Sigma}_{12}^{\mu} $. 
In contrast to the local density approximation 
where the self-energy is a function of the total Fermi momentum 
$k_{F_{tot}}$ the self-energy in CNM 
depends on the densities of the two subsystems and their 
streaming velocities \cite{sehn96,lca}. 

In the following we will only consider the symmetric case 
($k_{F_{1}}=k_{F_{2}}$). 
In Fig. 2 the equation-of-state's in symmetric 
colliding nuclear matter are shown for streaming velocities 
of the subsystem currents 
$u= |{\bf u}_i| =$ 0.2/0.4/0.6/0.8 (in units of $c$) which 
correspond to incident laboratory energies of 
$E_{\rm lab}$=0.08/0.36/1.05/3.34 GeV/nucleon. $u = 0$ is 
the isotropic case (ground state NM). The energy per particle 
is defined in the usual way as 
\beq
E_{12} (\varrho_{12},u) = {\rm T}^{00}_{12}/ \varrho_{12} -M
\label{ekin}
\eeq
with $ \varrho_{12} = \sqrt{ j_{12}^2} 
= < 1 >_{12} |_{c.m.}$ the invariant baryon density in the 
c.m. frame of the two currents, i.e. the frame where 
${\bf j}_{12} =0$. At high streaming velocities 
the ``EOS'' is significantly stiffer then in ground state 
nuclear matter because the energy per particle $E_{12}$ includes 
the contribution form the relative motion of the two 
currents. The general repulsive character of the momentum 
dependent part of the nuclear interaction leads to a strongly 
enhanced repulsive component which increases with 
an increasing amount of relative motion in the system.
\begin{figure}[h]
\unitlength1cm
\begin{picture}(8.,10.0)
\put(-0.5,0.0){\makebox{\epsfig{file=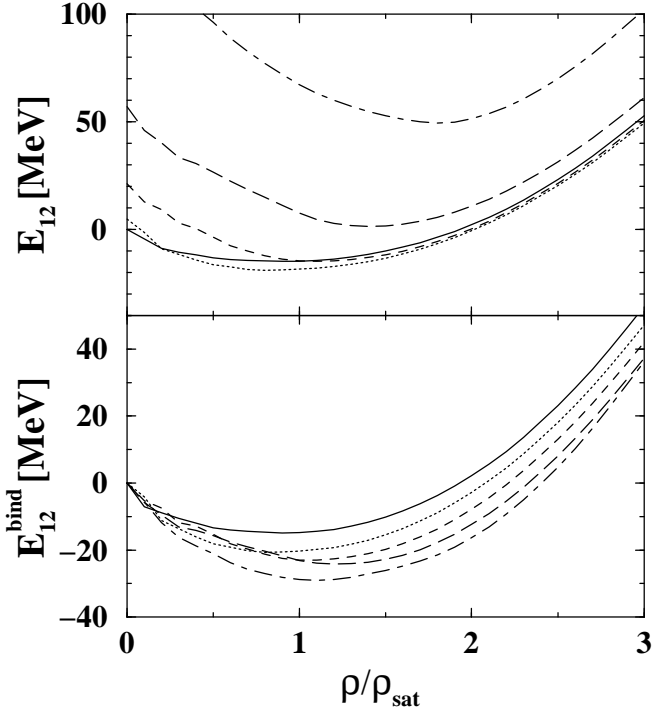,width=8.5cm}}}
\end{picture}
\caption{EOS in nuclear matter (solid) and colliding nuclear 
matter determined in the DBHF model. 
The upper part shows the total energy per particle 
$E_{12}$ as a function of the c.m. total density. The 
streaming velocities are $u$=0.2 (dotted), 0.4 (dashed), 
0.6 (long-dashed), 0.8 (dot-dashed). 
The lower part shows the effective EOS, i.e. the 
binding energy per particle $E_{12}^{\rm bind}$ where the 
kinetic energy of the relative motion in CNM has been 
subtracted. 
}
\label{Fig2}
\end{figure}

However, a meaningful discussion of non-equilibrium effects 
with respect to the ground state EOS should be based on 
the binding energy and thus the contribution from the relative 
motion of the two currents has to be subtracted. This leads to 
an effective EOS in colliding nuclear matter which is 
directly linked to the hydrodynamical picture \cite{stoecker86}. 
To do so, we subtract the kinetic energy of the relative motion 
with respect to single nuclear matter at rest
\beq
{\cal E}_{\rm rel}(\varrho_{12},u) = < E^* - M^* >_{12} 
-  < E^* - M^* >_{u=0}
\quad .
\eeq
The kinetic energy of a 
nucleon inside the medium $E^{*}_{\rm kin} =  E^* - M^* = 
\sqrt{ {\bf k}^{*2} +  M^{*2}} -  M^*$ calculated in CNM or NM, 
respectively, is thereby averaged of the corresponding configurations.  
Thus one obtains the binding energy per particle 
\beq
E_{12}^{\rm bind}(\varrho_{12},u) = \frac{ T^{00}_{12} 
-{\cal E}_{\rm rel}}{\varrho_{12}} -M
\label{ebind}
\eeq
as a function of the total c.m. density 
$\varrho_{12}$ and the c.m. streaming velocity $u$. 
The binding energy, 
i.e. the effective EOS in colliding nuclear matter, is shown in the 
lower part of Fig. 2. The effective EOS appears 
softer and even more attractive compared to ground state nuclear 
matter. This is a general feature of colliding nuclear 
matter and can be understood by a very transparent and model 
independent argument:

Let us consider two currents with 
sufficiently high $u$, i.e. well separated Fermi 
ellipsoids in momentum space ($\Theta_1 \cdot \Theta_2 =0$). 
For the following discussion we assume that the self-energies 
have no explicit momentum dependence, like in 
relativistic mean field (RMF) \cite{sw86} or density dependent 
RMF theory \cite{fule95}. 
Then the energy density (in the c.m. frame) is given by 
\beq
 T^{00}_{12}(\varrho_{12},u) = <E^* >_{12} - 
[ \Sigma_{S_{12}} \varrho_{s_{12}} 
+ \Sigma^{0}_{12} \varrho_{12} ]/2
\label{ebind2}
\eeq 
and the non-locality of the system, i.e. the 
high relative momenta of the separated ellipsoids enter only via 
the momentum dependence of $E^*$. The separation of projectile 
and target nucleons in momentum space 
increases the phase space volume since in both currents states 
are occupied up to $k_F = 0.5^{\frac{1}{3}} k_{F_{tot}}$. For 
a purely local interaction which is insensitive to the relative 
momenta of the currents this increase of phase space  
would have the same effect as the 
introduction of an additional virtual degree of freedom 
as illustrated in Fig.3. Thus, in a mean field approach 
the effective binding energy (\ref{ebind},\ref{ebind2}) can be 
approximated by a modification of 
the corresponding expression for single nuclear matter 
\beq
E_{12}^{\rm bind}(\varrho_{12},u) \simeq 
\left[ <E^* >_{1} - \frac{1}{2} \left( \Sigma_{S_{1}} \varrho_{s_{1}} 
+ \Sigma^{0}_{1} \varrho_{1} \right) \right] / \varrho_{1} -M
\quad .
\label{eos3}
\eeq

\begin{figure}[h]
\unitlength1cm
\begin{picture}(8.,6.0)
\put(-0.5,0.5){\makebox{\epsfig{file=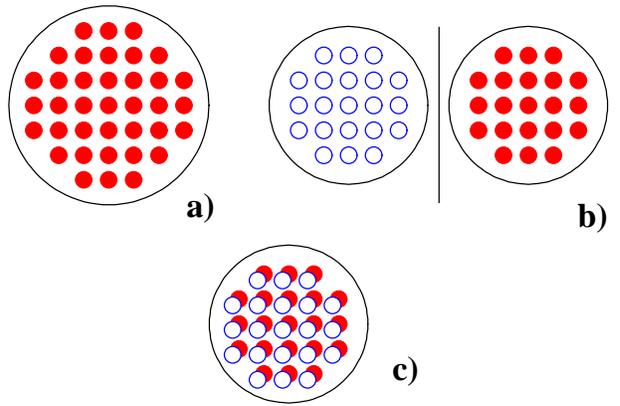,width=8.0cm}}}
\end{picture}
\caption{Schematic representation of the phase space in 
nuclear matter (a), colliding nuclear matter (b) and in 
colliding nuclear matter as experienced by a local potential 
(c).
}
\label{Fig3}
\end{figure}

To do so, on the right hand side of Eq. (\ref{eos3}) all quantities, i.e. 
$ \varrho_{1},\Sigma , \dots$  are obtained 
by the integration over one Fermi sphere at rest 
with the Fermi momentum $k_F$ corresponding to the 
density of only one current, however, taking thereby an 
enhanced phase space factor $\kappa=8$ in Eq. (\ref{confint}). For the 
vector density this leads to a linear dependence 
$ \varrho_{1}(k_F, \kappa=8) = 2\varrho_{1}(k_F, \kappa=4)$ which 
restores the total density. The dependence of the scalar 
density $\varrho_{s_{1}}$ is non-linear. 
However, the total c.m. vector density in colliding nuclear matter is 
still enhanced by a $\gamma$-factor, i.e. $\varrho_{12}(k_F,k_F,u)  
= \gamma (u) \varrho_{1}(k_F, \kappa=8)$ which does not completely 
cancel in the effective binding energy and leads to a stronger 
repulsion originating form the vector field 
as compared to the approximation (\ref{eos3}). 
In Fig.4 the corresponding EOS (\ref{eos3}) is shown as obtained in the 
density dependent RMF approach \cite{fule95} to the DBHF 
model. Here we neglect the explicit, but weak \cite{boelting99} 
momentum dependence of the DBHF self-energies. 
Varying in these calculations $\kappa$ from 4 to 8 illustrates 
the phase space effects. There is, of course, 
no exact agreement of the simplified ansatz of Eq. 
(\ref{eos3}) with full self-consistent CNM calculations (Fig.2), 
in particular at high densities, which is expected.  
With increasing density self-consistency effects between 
the effective mass $M^*$ and the configuration (\ref{conf}) itself 
become more important. A reduced in-medium mass 
$M^*$ leads to reduced momenta $k_{\mu}^* = u_\mu M^*$ at fixed 
velocities and to a shift of the Fermi ellipsoids. 
Thus at high densities Pauli blocking effects persist for  
interacting two-Fermi-ellipsoid configurations even at 
high relative velocities \cite{sehn96}. Furthermore, the relativistic 
mean field approximation is non-local through the momentum 
dependence via $E^*$ in (\ref{eos3}). 
However, it becomes clear from this comparison that 
the leading order effect for the EOS in colliding nuclear matter is 
the separation of phase space. 

\begin{figure}[h]
\unitlength1cm
\begin{picture}(8.,6.0)
\put(-1.5,0.0){\makebox{\epsfig{file=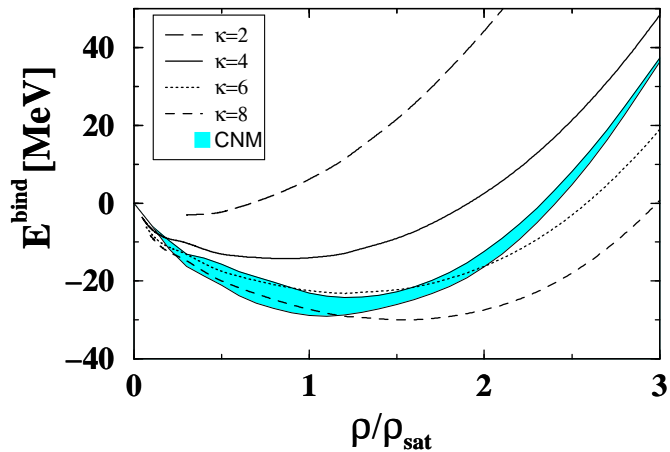,width=10.0cm}}}
\end{picture}
\caption{EOS in the relativistic density dependent 
mean field approximation to 
the DBHF model. The number of degrees of freedom is varied 
from $\kappa=2$ (neutron matter), $\kappa=4$ (nuclear matter) to 
$\kappa=6,8$. The shadowed band is limited by the EOSs in colliding 
nuclear matter with streaming velocities $u=0.6/0.8$ (upper/lower 
curves).
}
\label{Fig4}
\end{figure}

It is important to realize that this type of 
phase space effects is not included in standard transport 
calculations for heavy ion collisions, 
even when momentum dependent interactions are 
used. Phenomenological potentials \cite{welke88} 
\beq
U(\varrho,{\bf k} ) = U_{\rm loc}(\varrho ) + U_{\rm nonloc} (\varrho,{\bf k})
\label{pot}
\eeq
are usually composed by a local, density dependent potential 
$ U_{\rm loc}(\varrho )$ and 
a non-local momentum dependent part 
$U_{\rm nonloc} (\varrho,{\bf k}) 
= \int d^3 k' f({\bf k}') V({\bf k}-{\bf k}')$ 
with $V$ an effective two-body-interaction. In Hartree-Fock approximation 
these two terms correspond to the direct (Hartree) and the exchange (Fock) 
part of the potential. In ground state nuclear matter 
the Fock terms give usually small corrections 
to the EOS (mean field dominance) \cite{sw86}. In colliding 
nuclear matter the non-local part of the interaction is 
responsible for the strong repulsion seen in Fig.2. 
Applying potentials of the form (\ref{pot}) 
in transport calculations for heavy ion collisions the Fock part 
$U_{\rm nonloc}(\varrho,{\bf k})$ accounts by definition 
properly for the actual momentum 
space configurations $f({\bf k}')$, given e.g. by testparticle 
distributions. The mean field or local part is, however, 
decoupled from the anisotropy of the phase space since it is 
parameterized as a function of the total density. 
Consequently, $U_{\rm loc}(\varrho )$ reflects a density 
dependence which is only correct in equilibrated nuclear matter but does 
not apply to anisotropic momentum space configurations.

To summarise, the equation of state probed by the compression phase 
in energetic heavy ion reactions is to large extent governed by 
local non-equilibrium. The corresponding separation of phase space 
can be regarded as the introduction of an effectively new degree 
of freedom which lowers the binding energy per particle and 
makes the effective EOS seen in heavy ion reactions 
significantly softer. We conclude that this ``trivial'' but 
leading order effect should be taken into account 
when conclusions on the EOS are drawn from heavy ion collisions. \\
\begin{acknowledgments}
The auhtors thank H.H. Wolter, Amand Faessler and 
P. Danielewicz for fruitful discussions.
\end{acknowledgments}


\begin{thebibliography}{99}

\bibitem{stoecker86}
H. St\"ocker and W. Greiner, Phys. Rep. {\bf 137} (1986) 277.

\bibitem{ritter97}
W. Reisdorf and H.G. Ritter, 
Annu. Rev. Nucl. Part. Sci. {\bf 47} (1997) 663; 
N. Herrmann, J.P. Wessels, T. Wienold, 
Annu. Rev. Nucl. Part. Sci. {\bf 49} (1999) 581.


\bibitem{eflow}
P. Danielewicz, Roy A. Lacey, et al., 
Phys. Rev. Lett. {\bf 81} (1998) 2438; 
C. Pinkenburg {\it et al.}, Phys. Rev. Lett. {\bf 83} (1999) 1295.

\bibitem{AiKo85}
J. Aichelin and C.M. Ko, Phys. Rev. Lett. {\bf 55}, 2661 (1985).

\bibitem{sturm00}
C. Sturm {\it et al.}, 
Phys. Rev. Lett. {\bf 886} (2001) 39.

\bibitem{fuchs00}
C. Fuchs {\it et al.}, Phys. Rev. Lett. in press.

\bibitem{Cassing99}
W. Cassing and E.L. Bratkovskaya, Phys. Reports
{\bf 308} (1999) 65

\bibitem{Ai91} 
J. Aichelin, Phys. Reports {\bf 202} (1991) 233.

\bibitem{welke88}
G.M. Welke {\it et al.}, Phys. Rev. C {\bf 38} (1988) 2101; 
C. Gale {\it et al.}, Phys. Rev. C {\bf 41} (1990) 1545.

\bibitem{dani00}
P. Danielewicz, Nucl. Phys. {\bf A673} (2000) 275.

\bibitem{btm90}
   W. Botermans, R. Malfliet, 
   Phys. Rep. {\bf 198} (1990) 115.

\bibitem{gmat1}
   N. Ohtsuka, R. Linden,  A. Faessler, F.B. Malik,
   Nucl.Phys. {\bf A465} (1987) 550; 
   I. Izumoto, S. Krewald, A. Faessler, 
   Nucl.Phys. {\bf A341} (1980) 319.  

\bibitem{sehn96}
   L. Sehn, H.H. Wolter,  Nucl. Phys. {\bf A601} (1996) 473;
   C. Fuchs, L. Sehn, H.H. Wolter, Nucl. Phys. {\bf A601} (1996) 505; 
   L. Sehn, thesis (1990) unpublished.

\bibitem{lca}
C. Fuchs, T. Gaitanos, H. H.~Wolter, Phys. Lett. {\bf B381} (1996) 23; 
T. Gaitanos, C. Fuchs, H. H.~Wolter, Nucl. Phys. {\bf A650} (1999) 97.

\bibitem{rami99}
F. Rami et al. (FOPI Collaboration), Phys. Rev. Lett. {\bf 84} (2000) 1120.

\bibitem{essler97}
C. Fuchs, P. Essler, T. Gaitanos, H. H.~Wolter, 
Nucl. Phys. {\bf A626} (1997) 987.

\bibitem{dbhf}
   B. ter Haar, R. Malfliet, 
   Phys. Rep. {\bf 149} (1987) 207.

\bibitem{boelting99}
T. Gross-Boelting, C. Fuchs, and A. Faessler,
Nucl. Phys. {\bf A648} (1999) 105.

\bibitem{fopi95}
P. Dupieux {\it et al.}, Nucl. Phys. {\bf A587} (1995) 802.

\bibitem{hs86}
        B. D. Serot, J. D. Walecka, 
        Advances in  Nuclear Physics, {\bf 16}, 1,
        eds. J. W. Negele, E. Vogt, (Plenum, N.Y., 1986)

\bibitem{fule95}
        C. Fuchs, H. Lenske, H. H. Wolter,
        Phys. Rev. {\bf C52} (1995) 3043; 
H. Lenske and C. Fuchs, Phys. Lett. {\bf B345} (1995) 355.

\bibitem{sw86}
        B. D. Serot, J. D. Walecka, 
        Advances in  Nuclear Physics, {\bf 16}, 1,
        eds. J. W. Negele, E. Vogt, (Plenum, N.Y., 1986);\\
        Int. J. Mod. Phys. {\bf E6} (1997) 515.

\end{thebibliography}
\end{document}